\documentclass[prd,onecolumn,showpacs,superscriptaddress,nofootinbib,floatfix,11pt]{revtex4-2}

\usepackage{amsfonts}
\usepackage{amsmath}
\usepackage{amssymb}
\usepackage{array}
\usepackage{bm}
\usepackage{braket}
\usepackage{breqn}
\usepackage{color}
\usepackage{dcolumn}
\usepackage{epsfig}
\usepackage{epsf}
\usepackage{epstopdf}
\usepackage{float}
\usepackage{graphicx}
\usepackage{graphics}
\usepackage{harpoon}
\usepackage{indentfirst}
\usepackage{mathrsfs}
\usepackage{multirow}
\usepackage{slashed}
\usepackage{natbib}
\usepackage{makecell}
\usepackage{mathtools}
\usepackage{hyperref}
\usepackage{dsfont}
\usepackage{orcidlink}

\numberwithin{subsection}{section}

\newcommand{\beq}{\begin{eqnarray}}
\newcommand{\eeq}{\end{eqnarray}}

\newcommand{\mO}{\mathcal{O}}
\newcommand{\mM}{\mathcal{M}}
\newcommand{\md}{\mathrm{d}}

\newcommand{\itp}
{\affiliation{Institute of Theoretical Physics, Chinese Academy of Sciences, Beijing 100190, China}}

\newcommand{\hebtu}{\affiliation{Department of Physics and Hebei Key Laboratory of Photophysics Research and Application,\\
Hebei Normal University, Shijiazhuang 050024, China}}

\newcommand{\seu}{\affiliation{School of Physics, Southeast University, Nanjing 211189, China}}

\begin{document}

\title{
Axion-like particle production from lepton-nucleon scattering\\ in chiral effective theory
}

\author{Qian-Qian Guo\orcidlink{0009-0006-0520-5971}}
\email{qianqianguo@seu.edu.cn}
\hebtu\seu

\author{Xiong-Hui Cao\orcidlink{0000-0003-1365-7178}}
\email{xhcao@itp.ac.cn}
\itp

\author{Zhi-Hui Guo\orcidlink{0000-0003-0409-1504}}\email{zhguo@hebtu.edu.cn}
\hebtu

\author{Hai-Qing Zhou\orcidlink{0000-0002-1135-8951}}\email{zhouhq@seu.edu.cn}
\seu

\begin{abstract}
In this work we study the axion/axion-like particle production from the lepton-nucleon scattering in the low-energy region, i.e., the $\ell N\to \ell N a$ processes, $\ell$ being the electron or muon and $N$ the proton or neutron. We simultaneously include three different types of axion interaction couplings within the chiral effective field theory, namely the axion-nucleon-nucleon couplings $g_{aNN}$, axion-photon-photon coupling $g_{a\gamma\gamma}$ and axion-photon-vector meson resonances couplings $g_{\rho a\gamma}$ and $g_{\omega a\gamma}$. Vast inputs from the lattice QCD and hadron phenomenological studies are used to fix the unknown couplings. The relative strengths of different axion interactions in the $\ell N\to \ell N a$ processes are then revealed. We provide detailed predictions for the differential cross sections with respect to various angles and axion energy, as well as the total cross sections in the low-energy region around production thresholds, both for the Kim-Shifman-Vainstein-Zakharov (KSVZ) and Dine-Fischler-Srednicki-Zhitnitsky (DFSZ) axion models. 
\end{abstract}
\date{\today}
\maketitle

\section{Introduction}

Since the proposal of the axion in the seminal works~\cite{Peccei:1977hh,Peccei:1977ur,Weinberg:1977ma,Wilczek:1977pj}, this intriguing hypothetical particle has been the focus in many research branches of physics, including particle and nuclear physics, astronomy, cosmology, optics, atomic physics, etc.~\cite{Kim:2008hd,Graham:2015ouw,Irastorza:2018dyq,Sikivie:2020zpn,Choi:2020rgn,DiLuzio:2020wdo,Cong:2024qly,Jiang:2024boi}. As a natural extension of the axion proposal, the axion-like particle (ALP), whose mass and decay constant disentangle from the severely constrained relation of the axion case, has also attracted intensive attention. Particularly, the search for axion and ALP, which will be collectively denoted as ALP in this work for simplicity, constitutes a prominent subject at lepton fixed-target or beam-dump facilities, such as LDMX~\cite{Mans:2017vej,Akesson:2022vza}, M3~\cite{Kahn:2018cqs}, BDX~\cite{Izaguirre:2013uxa,Izaguirre:2014dua},  NA64~\cite{Banerjee:2019pds,NA64:2020qwq,NA64:2021xzo,Sieber:2021fue,NA64:2024klw,NA64:2024nwj}, MuSIC~\cite{Davoudiasl:2024fiz}, EIC~\cite{Balkin:2023gya}, EicC~\cite{Anderle:2021wcy,Gao:2024rgl} and so on. Due to their high luminosity, these kinds of experiments are expected to offer invaluable environments to impose strong constraints on the ALP parameters. 

The fundamental process in the lepton fixed-target or beam-dump experiments is the lepton-nucleon scattering, that is, $\ell N\to \ell N$. The ALP production process in these kinds of experiments is given by $\ell N\to \ell N a$. This latter amplitude has only been rarely studied in the literature within simple models, e.g., by only invoking the photon exchange mechanism between the lepton and the nucleon. In addition, most previous studies rely on the assumption of axion bremsstrahlung off the lepton or two-photon annihilation to ALP, i.e., focusing on the roles of the axion-lepton and axion-photon couplings in the $\ell N\to \ell N a$ processes~\cite{Tsai:1986tx,Darme:2020sjf,Liu:2023bby,Sieber:2023nkq,Gao:2024rgl,Ponten:2024grp}. 
As a novelty, we push forward in this work the investigation of the $\ell N\to \ell N a$ processes in two main aspects. 
First, chiral effective field theory (EFT) will be employed to calculate the $\ell N\to \ell N a$ amplitudes, by incorporating the axion-nucleon interaction vertex up to $O(p^2)$, which allows us to examine the roles of the axion-nucleon couplings and also provides a complementary study to previous works relying on the axion-lepton interaction. Second, in addition to the photon exchange, we will take into account the contributions from the light-flavor vector ($V$) resonances via the $Va\gamma$ interacting vertices, with $V=\rho$ and $\omega$, which have been demonstrated to be dominant in the axion photoproduction amplitude off the neutron target~\cite{Cao:2024cym} and $e^+ e^- \to \omega a$~\cite{Bai:2024lpq}. Furthermore, the influence of the different microscopic mechanisms on the cross sections and the distributions of various angles and energies will be explored in detail for the $e N \to e N a$ processes.

It is pointed out that our calculation in the chiral EFT is only valid in the low-energy region around the production thresholds for the $\ell N\to \ell N a$ processes, which cannot be extended to the high-energy regime far beyond the GeV scale as in many of the fixed-target or beam-dump experiments. Due to the complexity of the hadron amplitudes, the strict way to calculate the hadron-related processes ranging from its threshold to high-energy region is still challenging. In this situation, chiral EFT can offer a useful and model-independent tool to constrain the hadron amplitudes~\cite{Gasser:1983yg,Gasser:1984gg,Ecker:1988te}. The reliable low-energy $\ell N\to \ell N a$ amplitude calculated within the framework of chiral EFT hopefully will serve as a theoretical constraint for the ALP-related hadron amplitudes in the higher-energy regime.

The paper is structured as follows. In Sec.~\ref{setion:basic formulas}, we discuss the relevant axion interactions with nucleon, photon, and vector mesons in chiral EFT, and then calculate the $\ell N \to \ell N a$ amplitudes. Phenomenological discussions, including the treatment of kinematics and differential cross sections, are presented in Sec.~\ref{sec:pheno}. A short summary and conclusions are given in Sec.~\ref{sec:sum}. 

\section{Calculation of the $\ell N \to \ell N a$ amplitudes in chiral EFT}
\label{setion:basic formulas}

\subsection{ALP interactions with nucleon, photon and vector mesons}

The hypothetical axion corresponds to a pseudo-Nambu-Goldstone boson (pNGB), resulting from the spontaneous breaking of the global $U(1)$ Peccei-Quinn (PQ) symmetry~\cite{Peccei:1977hh,Peccei:1977ur,Weinberg:1977ma,Wilczek:1977pj}. In the low-energy region much below the electroweak breaking scale, the generic QCD Lagrangian involving light-flavor quarks, gluons and ALP as dynamical degrees of freedom is given by 
\begin{align}\label{eq.L1}
    \mathcal{L}_{\rm QCD}^{\mathrm{ALP}}=\mathcal{L}_{\mathrm{QCD},0}-\bar{q} \mathcal{M}_q q+\frac{1}{2}\partial_\mu a\partial^\mu a-\frac{1}{2}m^2_{a,0}a^2 +\frac{a}{f_a} \frac{\alpha_s}{8 \pi} \sum_{i=1}^{8} G_{\mu \nu,i} \tilde{G}^{\mu \nu}_{i}  +\frac{\partial^\mu a}{2 f_a} J_\mu^{\mathrm{PQ}}\,, 
\end{align}
where $\mathcal{L}_{\mathrm{QCD},0}$ is the standard QCD Lagrangian in the chiral limit; the light-flavor quark matrix is $\mathcal{M}_q={\rm diag}\left(m_u,m_d,m_s\right)$; $G_{\mu\nu,i}$ and $\tilde{G}_{\mu \nu,i}=\frac{1}{2} \epsilon_{\mu \nu \alpha \beta} G^{\alpha \beta}_{i}$ stand for the gluon field tensor and its dual, with $i$ the color indices, and $\alpha_s$ corresponds to the strong coupling constant of QCD. We use the convention $\epsilon_{0123}=+1$ for the Levi-Civita tensor $\epsilon_{\mu\nu\alpha\beta}$ throughout this work. The ALP decay constant is denoted by $f_a$, and the bare ALP mass $m_{a,0}$ can be considered as an additional $U(1)$ PQ symmetry-breaking term. The preexisting axion-quark interaction term is described by the PQ quark current,  
\begin{align}
J_\mu^{\mathrm{PQ}}= \bar{q} \gamma_\mu \gamma_5 \mathcal{X}_q q\ ,
\end{align}
where only the diagonal matrix $\mathcal{X}_q=\operatorname{diag}\left(X_u,X_d,X_s\right)$ in the flavor space will be assumed. 
The anomalous axion-gluon interaction, i.e., the $aG\tilde{G}$ term in Eq.~\eqref{eq.L1}, is usually deemed to be the model-independent part of the QCD axion interactions, due to its key role in solving the strong CP problem of QCD~\cite{Peccei:1977hh,Peccei:1977ur,Weinberg:1977ma,Wilczek:1977pj}, while the last term in Eq.~\eqref{eq.L1}, i.e., the preexisting axion-quark operator, is subject to the ultraviolet axion model constructions, which is then considered as the model-dependent axion interactions. For instance, two well-established benchmark axion models, viz. the Kim-Shifman-Vainstein-Zakharov (KSVZ)~\cite{Kim:1979if,Shifman:1979if} and Dine-Fischler-Srednicki-Zhitnitsky (DFSZ)~\cite{Dine:1981rt,Zhitnitsky:1980tq} models, yield markedly different predictions to the preexisting axion-quark couplings
\begin{align}
    \begin{aligned}
    X_q^{\mathrm{KSVZ}} & =0\ , \\
    X_{u}^{\mathrm{DFSZ}} &=\frac{1}{3} \sin ^2 \beta\ , \quad 
    X_{d, s}^{\mathrm{DFSZ}} &=\frac{1}{3} \cos ^2 \beta\ ,
    \end{aligned}
\end{align}
where $\tan\beta=v_u/v_d$ stands for the ratio of vacuum expectation values of the up and down types of Higgs doublets in the DFSZ axion model~\cite{Dine:1981rt,Zhitnitsky:1980tq}. 

Apart from the model-dependent part of the axion-quark coupling, model-independent contribution to such coupling will be also induced by the $aG\tilde{G}$ term in Eq.~\eqref{eq.L1}. One way to explicitly calculate such contribution is to perform the axial transformation of the quark fields $q \rightarrow \exp \left(i \gamma_5 \frac{a}{2 f_a} \mathcal{Q}_a\right) q $ with the constraint ${\rm Tr}(\mathcal{Q}_a) =1$. This procedure introduces several new terms to the original Lagrangian in Eq.~\eqref{eq.L1}, one of which exactly cancels with the $aG\tilde{G}$ in Eq.~\eqref{eq.L1}. The remaining modifications from such procedure are 
\begin{align}\label{eq.Ma}
&\mathcal{M}_q\to     \mathcal{M}_a=\exp \left(-i \frac{a}{2f_a} \mathcal{Q}_a\right) \mathcal{M}_q \exp \left(-i \frac{a}{2f_a} \mathcal{Q}_a\right) \, , \\ &  
J_\mu^{\mathrm{PQ}}\to J_\mu^a=J_\mu^{\mathrm{PQ}}- \bar{q} \gamma_\mu \gamma_5 \mathcal{Q}_a q\,. \label{eq.Ja}
\end{align}
After the axial quark transformation $q \rightarrow \exp \left(i \gamma_5 \frac{a}{2 f_a} \mathcal{Q}_a\right) q $, the original Lagrangian in Eq.~\eqref{eq.L1} becomes 
\begin{align}\label{eq.L2}
    \mathcal{L}_{\mathrm{QCD}}^{\mathrm{ALP},\prime}=\mathcal{L}_{\mathrm{QCD},0}+\frac{1}{2}\partial_\mu a\partial^\mu a-\frac{1}{2}m^2_{a,0}a^2 -\bar{q} \mathcal{M}_a q +\frac{\partial^\mu a}{2 f_a} J_\mu^a \,,
\end{align}
where the quantities $\mathcal{M}_a$ and $J_\mu^a$ are given in Eqs.~\eqref{eq.Ma} and \eqref{eq.Ja}, respectively. 
To proceed the calculation, an explicit realization of the matrix $\mathcal{Q}_a$ needs to be provided and a frequently used form proposed in Ref.~\cite{Georgi:1986df} is given by 
\begin{align}\label{eq.qa}
\mathcal{Q}_a=\frac{\mathcal{M}_q^{-1}}{\operatorname{Tr} \mathcal{M}_q^{-1}} = \frac{1}{1+z+w} \operatorname{diag}(1, z, w) \,, 
 \qquad \left(z=\frac{m_u}{m_d}, \, w=\frac{m_u}{m_s} \right) \,,
\end{align}
which renders the leading-order (LO) ALP-meson mass-mixing term vanishing. 
To acquire the ALP-hadron interactions, one needs to match the ALP-quark operators to the chiral EFT. A detailed discussion about this matching procedure for our purpose has been provided in Ref.~\cite{Cao:2024cym} and references therein. We do not repeat the derivations here, but directly give the results below for simplicity. The relevant chiral Lagrangians to our calculations up to $\mO(p^2)$ involving ALP and nucleon read
\begin{align}\label{eq.lagan1}
    \mathcal{L}_{a N}^{(1)}&=\bar{N}\left(i \slashed{D}-m_N+\frac{g_A}{2} \slashed{u}  \gamma_5+\frac{g_{0}^{i}}{2}  \slashed{\tilde{u}}_{i} \gamma_5\right)N\ ,\\
    \mathcal{L}_{a N}^{(2)}&=\bar{N}\left(\frac{c_6}{8 m_N} F_{\mu \nu}^{+}+\frac{c_7}{8 m_N} \operatorname{Tr}\left[F_{\mu \nu}^{+}\right]\right)\sigma^{\mu \nu} N \ , \label{eq.lagan2}
\end{align}
where the superscripts $(1)$ and $(2)$ denote the chiral orders, the covariant derivative is $D_\mu=\partial_\mu-ie Q_N A_\mu$, with $A_\mu$ the photon field and $Q_N={\rm diag}\{ 1,0\}$ the electric charge matrix of the nucleon doublet, $F^{+}_{\mu\nu} =2eQ_N\left(\partial_\mu A_\nu-\partial_\nu A_\mu\right)$, and the isotriplet and the isosinglet components of the axial-vector currents take the form $u_\mu=c_{u-d} \frac{\partial_\mu a}{f_a} \tau^3$ and $ \tilde{u}_{\mu,i=(u+d,s)}=c_{i=(u+d,s)} \frac{\partial_\mu a}{f_a} \mathds{1}$ in order, with
\begin{align}
    c_{u \pm d}=\frac{1}{2}\left(X_u \pm X_d-\frac{1 \pm z}{1+z+w}\right) \ , 
    \qquad c_s=X_s-\frac{w}{1+z+w}  \ .
\end{align}
To define the nonperturbative matrix elements $\Delta q$ via $\langle N |\bar{q}\gamma^\mu\gamma_5q | N\rangle = s^\mu \Delta q$, with $s^\mu$ the nucleon ($N$) spin vector~\cite{GrillidiCortona:2015jxo}, one can express
the isotriplet ($g_{A}$) and isosinglet ($g_0^{i=u+d,s}$) axial-vector nucleon couplings in 
terms of $\Delta q$ as $g_A=\Delta u-\Delta d,\, g_{0}^{u+d} =\Delta u+\Delta d,\,g_{0}^{s} =\Delta s$. 
From the Lagrangian in Eq.~\eqref{eq.lagan1}, the ALP-nucleon couplings can be acquired via $f_a g_{aNN}=g_A c_{u-d} \tau^3+g_0^i c_i \mathds{1}$, whose explicit expressions for proton and neutron are~\cite{GrillidiCortona:2015jxo,Vonk:2020zfh,Vonk:2021sit}
\begin{align}\label{eq.gap}
    f_a g_{app}&=-\frac{\Delta u+z \Delta d+w \Delta s}{1+z+w}+\Delta u X_u+\Delta d X_d+ \Delta s X_s\ ,\\
   f_a  g_{ann}&=-\frac{z \Delta u+\Delta d+w \Delta s}{1+z+w}+\Delta d X_u+\Delta u X_d+ \Delta s X_s  \ .\label{eq.gan}
\end{align}
According to the Flavour Lattice Averaging Group (FLAG)~\cite{FlavourLatticeAveragingGroupFLAG:2021npn}, the values of the relevant matrix elements are $\Delta u  =0.847(50),\, \Delta d =-0.407(34),\, \Delta s=-0.035(13)$. The FLAG average values for the light-quark masses~\cite{FlavourLatticeAveragingGroupFLAG:2021npn} will be also used to determine $z=0.485(19)$ and $w=0.025(1)$ appearing in Eq.~\eqref{eq.qa}. In the KSVZ case, i.e., $X_u=X_d=X_s=0$, the coupling strength of the ALP and proton is $f_a g_{app}=-0.430(36)$, which is much larger than the ALP-neutron one with $f_a g_{ann}=-0.002(30)$. Nevertheless, such a conclusion may be highly sensitive to the inclusion of the model-dependent parameters $X_{u,d,s}$. 

Regarding the low-energy constants (LECs) $c_6$ and $c_7$ in Eq.~\eqref{eq.lagan2}, their values can be determined by $c_6=\kappa_p-\kappa_n, c_7=\kappa_n$, where $\kappa_p$ and $\kappa_n$ stand for the anomalous magnetic moments of proton and neutron in order. Using the PDG~\cite{ParticleDataGroup:2024cfk} central values $\kappa_p=1.793$ and $\kappa_n=-1.913$, one can obtain $c_6=3.706$ and $ c_7=-1.913$.

For later convenience, we give the relevant Feynman rules for the $NN\gamma$ and $NNa$ vertices up to $\mO(p^2)$. 
For the $NN\gamma$ vertices, its LO and NLO Feynman rules take the form 
\begin{align}\label{eq.vgnn}
v^\mu_{\gamma NN,\, \text{LO}}=ieQ_N\gamma^\mu\,, \qquad v^\mu_{\gamma NN,\, \text{NLO}}=-\frac{e\left(c_6 Q_N+c_7\right)}{2m_N}\sigma^{\mu\nu}q_\nu\,,
\end{align}
where $q$ corresponds to the incoming photon momentum. The LO Feynman rule for the $NNa$ vertex reads 
\begin{align}\label{eq.vann}
v^{\text{LO}}_{aNN}=-\frac{g_{aNN}}{2}\slashed{q}\gamma_5\,,
\end{align}
with $q$ the outgoing ALP momentum. The ALP-nucleon couplings $g_{aNN}$ are defined in Eqs.~\eqref{eq.gap} and \eqref{eq.gan}. It is noted that we have reshuffled the factor of $f_a$ into the definition of the $g_{aNN}$ couplings, while the $f_a$ factor is explicitly introduced in the vertex in Ref.~\cite{Cao:2024cym}. We note that the $NNa$ interacting vertex does not appear at $\mO(p^2)$ in the chiral EFT.

The anomalous ALP-photon interaction Lagrangian reads  
\begin{align}\label{eq.vagg}
\mathcal{L}_{a\gamma\gamma}=\frac{a}{f_a}  \frac{\alpha }{8 \pi}C_{a \gamma}F^{\mu\nu} \tilde{F}_{\mu\nu} \equiv  g_{a\gamma\gamma}\, a F^{\mu\nu} \tilde{F}_{\mu\nu} \,,
\end{align}
where $\alpha$ is the electromagnetic fine structure constant, and $F_{\mu\nu}$ and $\tilde{F}_{\mu \nu}$ denote the photon field strength tensor and its dual. The coupling $C_{a \gamma}$ receives both model-dependent and -independent pieces. In a recent $SU(2)$ chiral EFT calculation up to NLO, the value of the ALP-photon coupling is determined to be $C_{a\gamma}=\frac{E}{N}-1.92(4)$~\cite{GrillidiCortona:2015jxo}, where $N$ and $E$ are the model-dependent QCD and QED anomaly coefficients, and the number $-1.92(4)$ corresponds to the model-independent part. The $SU(3)$ NLO calculation predicts the model-independent part as $-2.05(3)$~\cite{Lu:2020rhp}. Recently, the model-independent part of the ALP-photon coupling is also calculated in $U(3)$ chiral EFT with the prediction of $-1.89(2)$, after including the linear isospin-breaking effect~\cite{Gao:2022xqz,Gao:2024vkw}. 

In Ref.~\cite{Cao:2024cym}, it was demonstrated that the $Va\gamma$ types of interacting vertices, with $V$ the light-flavor vector mesons $\rho$ and $\omega$, can play important roles in the $\gamma N\to a N$ processes, especially for the neutron channel. Therefore, one would expect that these kinds of interactions will also manifest their roles in the $e N \to e N a$ processes. 
The evaluation of the $Va\gamma$ coupling has been given in detail in Ref.~\cite{Cao:2024cym}, and we recapitulate the derivation to set up the notations. The $Va\gamma$ interacting vertex takes the form 
\begin{align}\label{eq.vvag}
\mathcal{M}_{Va \gamma}= e g_{Va\gamma} \epsilon_{\alpha \beta \rho \sigma} \epsilon_V^\alpha(q) \epsilon_\gamma^{*\beta}(k) q^\rho k^\sigma \ ,
\end{align}
where the $Va\gamma$ coupling is estimated using $g_{\small Va\gamma}=\sum_{(P=\pi^0,\eta,\eta')} \theta_{aP} g_{\small VP\gamma}$, with $\theta_{aP}$ the mixing factor between ALP and the light pseudoscalar $P$, and $g_{\small VP\gamma}$ denotes the $VP\gamma$ coupling. We will adopt the mixing forms $\theta_{aP}$ from Ref.~\cite{Alves:2024dpa} in our calculation. For the $VP\gamma$ couplings, we take the results from the resonance chiral theory in Refs.~\cite{Ruiz-Femenia:2003jdx,Chen:2012vw,Chen:2014yta,Yan:2023nqz}. For definiteness, we adopt the parameter values from Table I of Ref.~\cite{Yan:2023nqz}, which are determined through a joint fit to a large amount of experimental data, including various processes of $P\to V\gamma$, $V\to P\gamma^{(*)}$, $P \to \gamma \gamma^{(*)}$ as well as transitions involving $J/\psi$ and $\psi'$, i.e., the Set-I scenario of Ref.~\cite{Cao:2024cym}  will be employed in our analysis. The Set-II inputs in the former reference give roughly similar results, and we will not explicitly show such analysis here. For the explicit expressions of $\theta_{aP}$ and $g_{\small VP\gamma}$, we refer to Ref.~\cite{Cao:2024cym} for further details. 
In the QCD axion scenario, i.e., by setting $X_u=X_d=X_s=0$, the model-independent parts for the couplings of $g_{\rho^0 a\gamma}$ and $g_{\omega a\gamma}$, by taking the hadronic inputs from Ref.~\cite{Yan:2023nqz}, are given by $f_a g_{\rho^0 a\gamma}=-0.132(20)$ and $f_a g_{\omega a\gamma}=-0.077(7)$ in the case of $m_a=0$. 

In order to include the $Va\gamma$ interaction in the $e N \to e N a$ amplitudes, one also needs the $VNN$ interacting vertices as additional inputs. Following the phenomenological discussions in Refs.~\cite{Bauer:2012pv,Janssen:1996kx,Ronchen:2012eg}, we take the $VNN$ interacting Lagrangian 
\begin{align}\label{eq.L_rhoNN}
    \mathcal{L}_{VNN}=\Bar{N}\bigg( g_{\rho NN}\vec{\rho}_\mu\cdot\vec{\tau} +g_{\omega NN}\omega_\mu \bigg)\gamma^\mu N+\frac{G_\rho}{2}\Bar{N}\vec{\rho}_{\mu\nu}\cdot \vec{\tau}\sigma^{\mu\nu}N\ ,
\end{align}
where $\vec{\tau}$ denotes the Pauli matrices and $\vec{\rho}_{\mu \nu}=\partial_\mu \vec{\rho}_\nu-\partial_\nu \vec{\rho}_\mu$. For the values of the $VNN$ couplings, the determinations of $g_{\rho NN}=3.25$, $g_{\omega NN}=11.7$, and $G_\rho=-10.6~\mathrm{GeV}^{-1}$ from the J\" ulich model~\cite{Janssen:1996kx,Ronchen:2012eg} will be used in our analysis. The $\omega$ tensor coupling is conventionally set to zero in meson-nucleon reaction studies~\cite{Janssen:1996kx,Ronchen:2012eg}.

\subsection{Calculation of the $\ell N\to \ell N a$ amplitudes}

\begin{figure}[htbp]
\centering
\includegraphics[height=8cm]{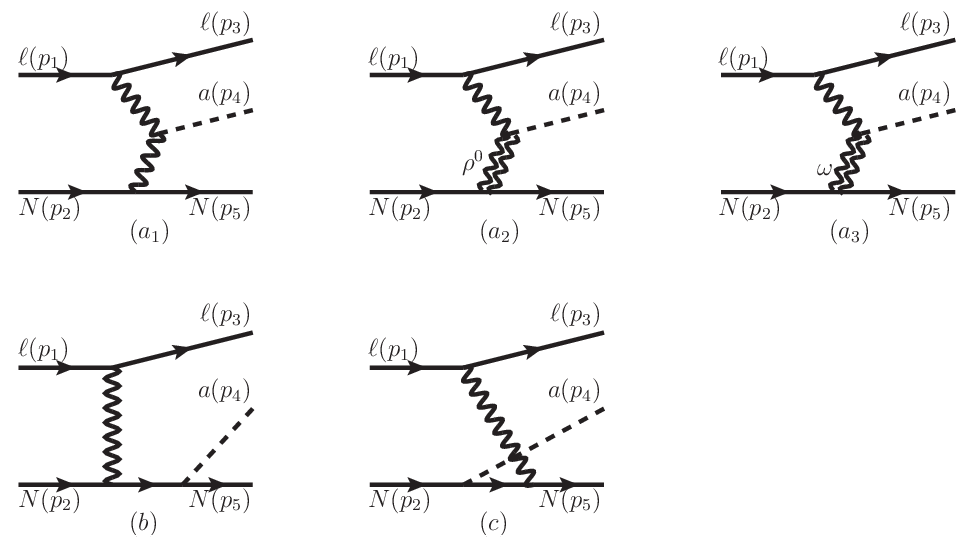}
\caption{The Feynman diagrams of $\ell N\rightarrow \ell N a$.}
\label{Figure:ep-epa-tree}
\end{figure}

With the above preparations, we are ready to calculate the $\ell (p_1) N(p_2)\to \ell(p_3) N(p_5) a(p_4)$ amplitudes, and the corresponding Feynman diagrams are shown in Fig.~\ref{Figure:ep-epa-tree}.

By taking the interacting vertices in Eqs.~\eqref{eq.vgnn}-\eqref{eq.vvag}, one can calculate the amplitudes corresponding to the Feynman diagrams in Fig.~\ref{Figure:ep-epa-tree}; the explicit expressions are given by 
\begin{eqnarray}
\mathcal{M}_{a_1} &=& -i\bar{u}(p_3,m_{\ell})\Gamma_{\gamma \ell\ell}^{\mu} u(p_1,m_{\ell}) \bar{u}(p_5,m_N)\Gamma_{\gamma NN}^{\sigma}(p_5-p_2) u(p_2,m_N) \Gamma_{a \gamma\gamma}^{\nu\eta}(p_1-p_3,p_5-p_2) \nonumber\\
~~~~~~&& \times S_{\gamma,\mu\nu}(p_1-p_3)S_{\gamma,\eta\sigma}(p_5-p_2) \ ,\nonumber\\
\mathcal{M}_{a_2} &=& -i\bar{u}(p_3,m_{\ell})\Gamma_{\gamma \ell\ell}^{\mu} u(p_1,m_{\ell}) \bar{u}(p_5,m_N)\Gamma_{\rho^{0} NN}^{\sigma}(p_5-p_2) u(p_2,m_N) \Gamma_{\rho^{0} a\gamma}^{\nu\eta}(p_1-p_3,p_5-p_2) \nonumber\\
~~~~~~&& \times S_{\gamma,\mu\nu}(p_1-p_3)S_{(\rho^{0}),\eta\sigma}(p_5-p_2) \ , \nonumber\\
\mathcal{M}_{a_3} &=& -i\bar{u}(p_3,m_{\ell})\Gamma_{\gamma \ell\ell}^{\mu} u(p_1,m_{\ell}) \bar{u}(p_5,m_N)\Gamma_{\omega NN}^{\sigma} u(p_2,m_N) \Gamma_{\omega a\gamma}^{\nu\eta}(p_1-p_3,p_5-p_2) \ ,\nonumber\\
~~~~~~&& \times S_{\gamma,\mu\nu}(p_1-p_3)S_{(\omega),\eta\sigma}(p_5-p_2)\ , \nonumber\\
\mathcal{M}_{b} &=& -i\bar{u}(p_3,m_{\ell})\Gamma_{\gamma \ell\ell}^{\mu} u(p_1,m_{\ell}) \bar{u}(p_5,m_N)\Gamma_{aNN}(p_4) S_N(p_1+p_2-p_3) \Gamma_{\gamma NN}^{\nu}(p_1-p_3) u(p_2,m_N) \nonumber\\
~~~~~~&& \times S_{\gamma,\mu\nu}(p_1-p_3) \ ,\nonumber\\
\mathcal{M}_{c} &=& -i\bar{u}(p_3,m_{\ell})\Gamma_{\gamma \ell\ell}^{\mu} u(p_1,m_{\ell}) \bar{u}(p_5,m_N)\Gamma_{\gamma NN}^{\nu}(p_1-p_3) S_N(p_5-p_1+p_3) \Gamma_{aNN}(p_4) u(p_2,m_N) \nonumber\\
~~~~~~&& \times S_{\gamma,\mu\nu}(p_1-p_3) \ ,
\end{eqnarray}
where 
\begin{eqnarray}
\Gamma_{a\gamma\gamma}^{\mu\eta}(p_i,p_f) &=& 4i  {g_{a\gamma\gamma}} \varepsilon^{\mu\nu\eta\sigma}p_{i,\nu}p_{f,\sigma}, \nonumber\\
\Gamma_{aNN}(q) &=&  -\frac{g_{aNN}}{2} \slashed{q} \gamma^{5}, \qquad \Gamma_{\gamma \ell\ell}^{\mu} = -ie\gamma^{\mu}, \nonumber\\
\Gamma_{\gamma NN}^{\mu}(q) &=&  ie \left[Q_N \gamma^{\mu}-\frac{(c_6Q_N +c_7)}{4m_N}(\gamma^{\mu} \slashed{q}-\slashed{q}\gamma^{\mu})\right],  \nonumber\\
\Gamma_{\rho^{0} NN}^{\mu}(q) &=& i\left[ g_{\rho NN} \gamma^{\mu} - \frac{G_\rho}{2}(\gamma^{\mu} \slashed{q}-\slashed{q}\gamma^{\mu})\right],  \nonumber\\
\Gamma_{\omega NN}^{\mu} &=& i {g_{\omega NN}} \gamma^{\mu}, \quad \Gamma_{\rho^{0}(\omega)a\gamma}^{\mu\eta}(p_i,p_f) = ie  {g_{\rho^{0}(\omega)a\gamma}} \varepsilon^{\mu\nu\eta\sigma}p_{i,\nu}p_{f,\sigma}, \nonumber\\
S_{\gamma}^{\mu\nu}(k) &=& \frac{-ig^{\mu\nu}}{k^2+i\epsilon}, \quad S_{N}(k) = \frac{i(\slashed{k}+m_N)}{k^2-m_{N}^2+i\epsilon}, \nonumber\\
S_{(\rho^{0})}^{\mu\nu}(k) &=& \frac{-i\left(g^{\mu\nu}-\frac{k^{\mu}k^{\nu}}{m_{\rho^0}^2}\right)}{k^2-m_{\rho^{0}}^2+i\epsilon}, \quad S_{(\omega)}^{\mu\nu}(k) = \frac{-i\left(g^{\mu\nu}-\frac{k^{\mu}k^{\nu}}{m_{\omega}^2}\right)}{k^2-m_{\omega}^2+i\epsilon}   \,.
\label{vertexes}
\end{eqnarray}
The values of the various couplings, including $g_{a\gamma\gamma}, g_{aNN}, g_{\rho(\omega) NN}, g_{\rho(\omega) a\gamma}$, have been discussed previously. The complete $\ell N\to \ell N a$ amplitude is given by the sum of the individual terms 
\begin{equation}\label{eq.ampful}
 \mathcal{M}= \mathcal{M}_{a1}+\mathcal{M}_{a2}+\mathcal{M}_{a3}+\mathcal{M}_{b}+\mathcal{M}_{c}\,, 
\end{equation}
where $\mathcal{M}_{a1}$ represents the amplitude from the photon exchange, $\mathcal{M}_{a2}$ and $\mathcal{M}_{a3}$ are the results from the exchanges of $\rho$ and $\omega$ in order, and $\mathcal{M}_{b}, \mathcal{M}_{c}$ denote the amplitudes from the nucleon exchanges via the direct $aNN$ vertex.   

To end this section, we briefly discuss the connection between the lepton-nucleon and the lepton-nucleus amplitudes, since many of the fixed-target or beam-dump facilities employ nucleus targets, with the numbers of protons $Z_p$ and neutrons $Z_n$. When assuming the photon exchange as the only contribution and further keeping the LO $\gamma NN$ vertex in Eq.~\eqref{eq.vgnn}, the latter of which gives a vanishing coupling for the neutron, one can obtain $\mM_{l-{\rm nucleus}}^{\gamma-{\rm exchange,LO}} = Z_p \mM_{l-{\rm proton}}^{\gamma-{\rm exchange,LO}}$ within the coherent picture, which basically sum the contributions from individual protons inside the nucleus at the amplitude level. In this way, to take the amplitude squared $|\mM_{l-{\rm nucleus}}^{ \gamma-{\rm exchange,LO}}|^2$ as input will lead to the typical enhancement factor $Z_p^2$ in the lepton-nucleus cross section, which is exactly the formalism used in many previous studies\cite{Tsai:1986tx,Darme:2020sjf,Liu:2023bby,Sieber:2023nkq,Gao:2024rgl,Ponten:2024grp}. Nevertheless, our phenomenological studies in the next section show that the contributions from the exchanges of nucleons and vector mesons can be more profound than those of the photon exchange, at least in the low-energy region not far from the production thresholds. A reliable estimation of the cross sections needs to include the former two kinds of contributions, rather than purely the photon exchange. Furthermore, by including the NLO vertex in Eq.~\eqref{eq.vgnn} and also the exchanges of the vector mesons and nucleon, the amplitude of $\ell n\to \ell n a$ clearly does not vanish. Thus the more complete lepton-nucleus cross sections should also include the effects from the neutrons inside the nucleus, apart from those of the protons.   
Following the coherent picture, a naive way to construct the full lepton-nucleus amplitude would be $\mM_{\ell-{\rm nucleus}} = Z_p \mM_{\ell-{\rm proton}}^{\rm total}+  Z_n \mM_{\ell-{\rm neutron}}^{\rm total}$, where the total lepton-nucleon amplitudes $\mM_{\ell-{\rm proton}}^{\rm total}$ and $\mM_{\ell-{\rm neutron}}^{\rm total}$ should contain all the contributions from the three types of exchanges of photon, nucleon, and vector mesons in Eq.~\eqref{eq.ampful}.  
However, when summing the spins for the interference term of $Z_pZ_n(\mM_{\ell-{\rm proton}}^{\rm total}\mM_{\ell-{\rm neutron}}^{\rm total,*}+\mM_{\ell-{\rm proton}}^{\rm total}\mM_{\ell-{\rm neutron}}^{\rm total,*})$ arising from amplitude squared in the calculation of the lepton-nucleus cross sections, the formation details of the nucleus, such as the spin states of the protons and nucleons inside the nucleus, will generally be needed. Inclusion of such effect in the lepton-nucleus cross section clearly deserves an independent work which is beyond the scope of the present one. For simplicity, we will simply neglect the interferences of the proton and neutron amplitudes. In this case, the lepton-nucleus cross sections will be given by $\sigma_{\ell-{\rm nucleus}} = Z_p^2 \sigma_{\ell-{\rm proton}}+  Z_n^2\sigma_{\ell-{\rm neutron}}$, where the cross sections of $\sigma_{\ell-{\rm proton}}/\sigma_{\ell-{\rm neutron}}$ are determined by the total amplitudes $\mM_{\ell-{\rm proton}}^{\rm total}/\mM_{\ell-{\rm neutron}}^{\rm total}$, respectively.

\section{Phenomenological discussions}\label{sec:pheno}

\subsection{Phase space for the $\ell N\to \ell N a$ process}
\label{section:phasespace}

\begin{figure}[t]
\centering
\includegraphics[width=0.75\textwidth]{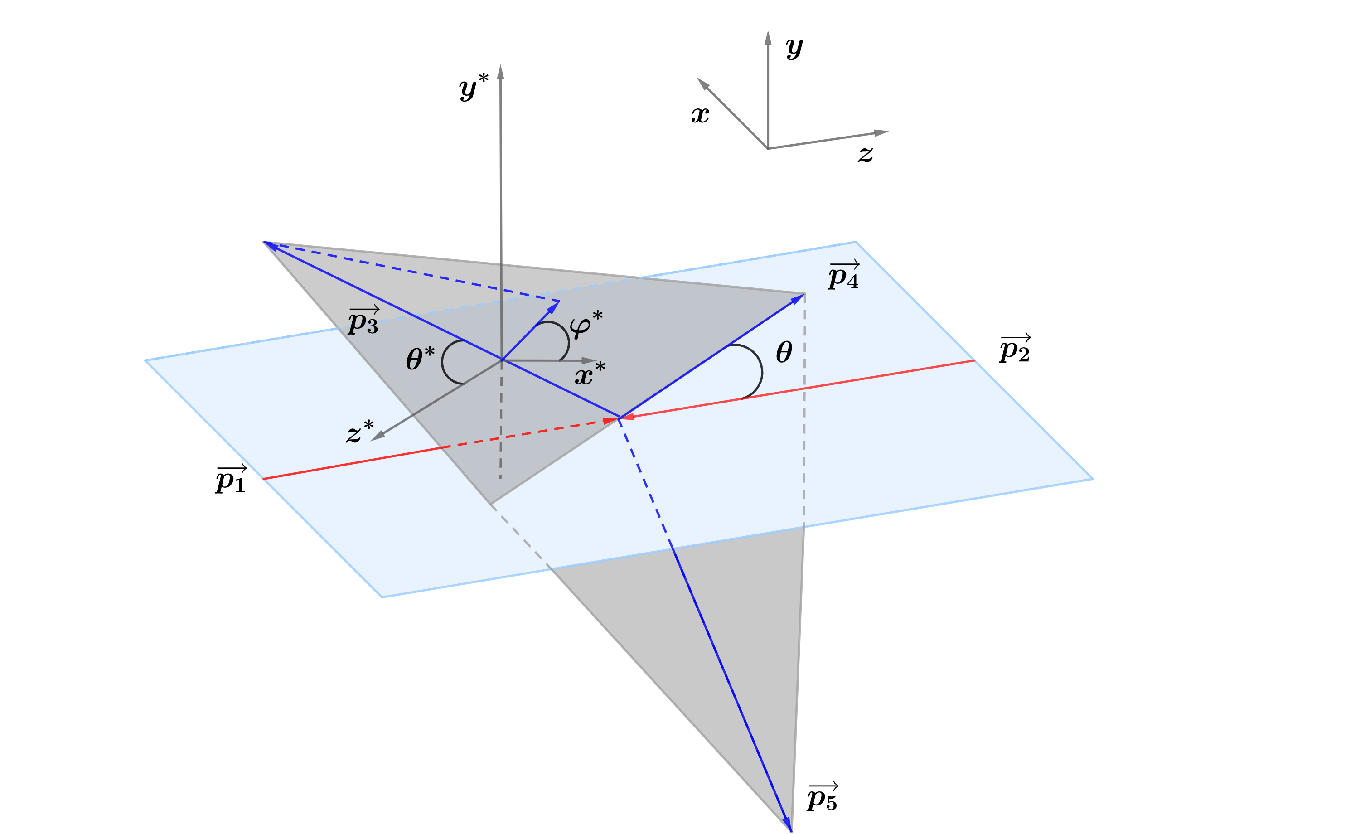}
\caption{Illustration of the kinematics in the CM frame of the incoming particles for the $\ell (p_1) N(p_2)\to \ell(p_3) N(p_5) a(p_4)$ reaction.}
\label{Figure:ep-epa-tree-CM}
\end{figure}

Before the phenomenological discussions, we first briefly discuss the way to calculate the phase space of the $2\to 3$ scattering. For the $\ell (p_1) N(p_2)\to \ell(p_3) N(p_5) a(p_4)$ process, it is convenient to work in the center of mass (CM) frame of the incoming lepton $(p_1)$ and the nucleon $(p_2)$. We illustrate the adopted kinematics in Fig.~\ref{Figure:ep-epa-tree-CM}. For the quantity defined in the CM frame of the outgoing lepton ($p_3$) and nucleon ($p_5$), we will introduce an asterisk to distinguish them from those in the CM frame of the incoming particles.  

The quantity $\theta$ corresponds to the scattering angle between the three momenta of $\vec{p}_1$ and $\vec{p}_4$ in the CM frame of the incoming particles. 
We use $\theta^\ast$ and $\varphi^\ast$ to denote the polar and azimuthal angles, respectively, of the outgoing lepton with $p_3$, in the CM frame of the outgoing lepton and nucleon with $p_3$ and $p_5$. 
With these notations, the momenta in the two different CM frames can be written as 
\begin{eqnarray}\label{eq.defpi}
p_{1\text{c}} &=& \left(E_1,0,0,\sqrt{E_1^2-m_1^2} \right)\,,  \nonumber \\
p_{2\text{c}} &=& \left(E_2,0,0,-\sqrt{E_2^2-m_2^2} \right)\,,  \nonumber \\
p_{3\text{c}}^\ast &=& (\frac{m_{35}^2+m_3^2-m_5^2}{2m_{35}},q_{35}^{\text{CM}}\sin\theta^\ast\cos\varphi^\ast,q_{35}^{\text{CM}}\sin\theta^\ast\sin\varphi^\ast,q_{35}^{\text{CM}}\cos\theta^\ast)\,,  \nonumber \\
p_{5\text{c}}^\ast &=& (\frac{m_{35}^2+m_5^2-m_3^2}{2m_{35}},-q_{35}^{\text{CM}}\sin\theta^\ast\cos\varphi^\ast,-q_{35}^{\text{CM}}\sin\theta^\ast\sin\varphi^\ast,-q_{35}^{\text{CM}}\cos\theta^\ast)\,,  
\end{eqnarray}
where $m_i$ correspond to the masses of the particles with momenta $p_i$, and other relevant quantities are given by 
\begin{align}
q_{35}^{\text{CM}} & =\frac{\lambda(m_{35},m_3,m_5)}{2m_{35}}\ ,\quad 
m_{35} =\sqrt{(p_3+p_5)^2} = \sqrt{s+m_4^2-2\sqrt{s}E_4}\,, \nonumber \\
s &=(p_1+p_2)^2\ ,\quad \lambda(x,y,z) = \sqrt{x^4+y^4+z^4-2x^2y^2-2x^2z^2-2y^2z^2}\,.
\end{align}

From the expressions of $p^\ast_{3c}$ and $p^\ast_{5c}$ in Eq.~\eqref{eq.defpi}, one can determine the corresponding expressions of $p_{3}$ and $p_{5}$ in the CM frame of the incoming particles by performing the Lorentz boost and rotation through  
\begin{eqnarray}
p_{3\text{c}} &=& \text{Rot}_{\text{y}}(\theta+\pi)\cdot \text{Boost}_{\text{z}}(\beta)\cdot p_{3\text{c}}^\ast \,, \nonumber \\
p_{5\text{c}} &=& \text{Rot}_{\text{y}}(\theta+\pi)\cdot \text{Boost}_{\text{z}}(\beta)\cdot p_{5\text{c}}^\ast \,, 
\end{eqnarray}
with
\begin{eqnarray}
\text{Boost}_{\text{z}}(\beta) &=&
\begin{bmatrix}
\gamma & 0 & 0 & \gamma\beta \\
0 & 1 & 0 & 0 \\
0 & 0 & 1 & 0 \\
\gamma\beta & 0 & 0 & \gamma 
\end{bmatrix}, \quad 
\text{Rot}_{\text{y}}(\theta) =
\begin{bmatrix}
1 & 0 & 0 & 0 \\
0 & \cos\theta & 0 & \sin\theta \\
0 & 0 & 1 & 0 \\
0 & -\sin\theta & 0 & \cos\theta 
\end{bmatrix}\,, 
\end{eqnarray}
and 
\begin{eqnarray}
\beta &=& \sqrt{1-\frac{1}{\gamma^2}}, \qquad \gamma = \frac{E_{35}}{m_{35}}, \qquad  E_{35} = \frac{s+m_{35}^2-m_4^2}{2\sqrt{s}}\,.
\end{eqnarray}

The momentum of the axion in the CM frame of the incoming particles is simply given by 
\begin{eqnarray}\label{eq.defp4}
p_{4\text{c}}= p_{1\text{c}}+p_{2\text{c}}-p_{3\text{c}}-p_{5\text{c}}\equiv(E_4,\vec{p}_4) \,.  
\end{eqnarray}

Following the standard convention for the calculation of the $2\to 3$ phase space~\cite{ParticleDataGroup:2024cfk}, after integrating out the azimuthal angle of $\varphi$ in the CM frame of the incoming particles, the cross section of the $\ell(p_1) N(p_2)\to \ell(p_3) N(p_5) a(p_4)$ can be written as 
\begin{eqnarray}\label{eq.defcs}
\sigma &=&\frac{1}{64(2\pi)^4}\frac{1}{\sqrt{(p_1\cdot p_2)^2-m_e^2m_p^2}}\int_{m_a}^{E_a^{\text{max}}}dE_a\sqrt{E_a^2-m_a^2}\int_{0}^{\pi}d\theta\sin\theta \nonumber \\
~~~~~~&& \times \int_{0}^{\pi}d\theta^{\ast}\sin\theta^{\ast}\int_{0}^{2\pi}d\varphi^{\ast}\frac{\lambda(m_{35},m_3,m_5)}{m_{35}^2}|\overline{\mathcal{M}}|^2\,,
\end{eqnarray}
where the amplitude squared $|\overline{\mathcal{M}}|^2$ has been evaluated with the spin average/sum of the initial/final states. Notice that we have used $E_a$ and $m_a$ to replace the notations $E_4$ and $m_4$ in Eq.\eqref{eq.defcs}, and the maximum value of the former quantity in the kinematically allowed region is 
\begin{eqnarray}\label{eq.p4max}
|\vec{p}_{a}|^{\text{max}} =\frac{\lambda(\sqrt{s},m_3+m_5,m_a)}{2\sqrt{s}}, \qquad E_a^{\text{max}} = \sqrt{(|\vec{p}_{a}|^{\text{max}})^2+m_a^2}\,. 
\end{eqnarray}
The relations in Eqs.~\eqref{eq.defpi}-\eqref{eq.defp4} are used to express $|\overline{\mathcal{M}}|^2$ in terms of the integration variables of $E_a$, $\theta$, $\theta^\ast$, and $\varphi^\ast$ in Eq.~\eqref{eq.defcs}. 

In the following discussions, we will separately study the impacts of the nucleon-, photon- and vector-meson-exchanges on the (differential) cross sections, so that one can clearly discern the relative importance of different contributions.

\subsection{Differential cross sections with respect to various angles and axion energy} 
\label{setion:dcs}

In this section, we investigate the differential cross sections as functions of various angles as mentioned above and also the axion energy. Since the chiral EFT is expected to be only reliable in the low-energy region, not far away from the production threshold, we fix the CM energy of the incoming $\ell N$ at $\sqrt{s} =1.2$ $\textrm{GeV}$ when studying the differential cross sections. 

\begin{figure}[t]
\centering
\includegraphics[width=0.75\textwidth]{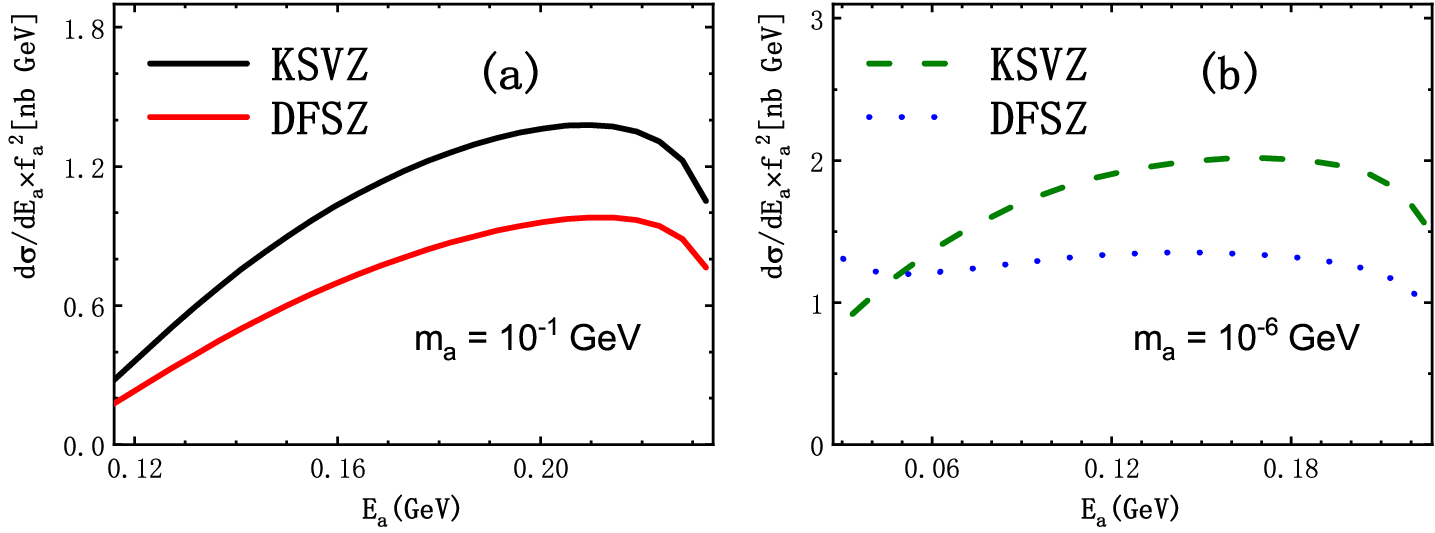}
\caption{Differential cross sections evaluated with the full amplitudes as a function of axion energy $E_{a}$ for the $ep\rightarrow epa$ process at $m_a=0.1$~GeV (a) and $10^{-6}$~GeV (b). }
\label{Figure:tcepEa}
\end{figure}

\begin{figure}[t]
\centering
\includegraphics[width=1\textwidth]{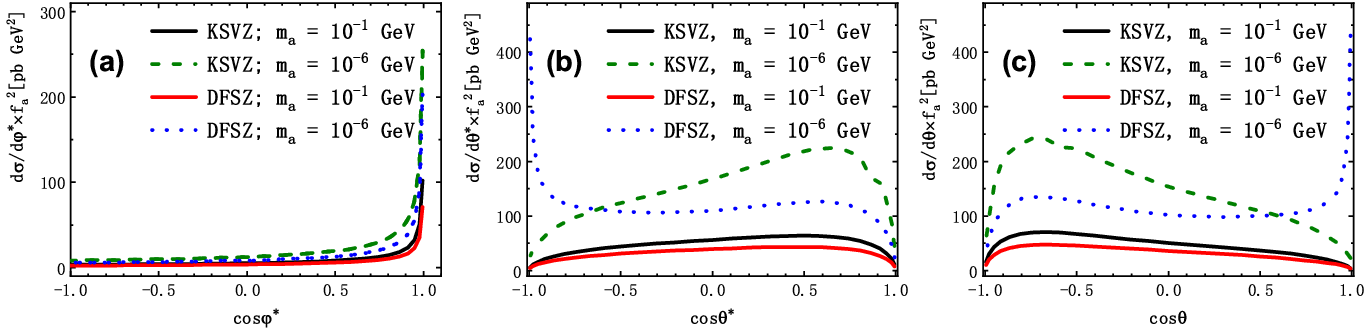}
\caption{Differential cross sections $\md\sigma/\md\varphi^*$ (a), $\md\sigma/\md\theta^*$ (b), and $\md\sigma/\md\theta$ (c) for $ep\rightarrow epa$, evaluated with full amplitudes and plotted against the cosine of the corresponding angle.}
\label{Figure:dcepangle}
\end{figure}

\begin{figure}[t]
\centering
\includegraphics[width=0.75\textwidth]{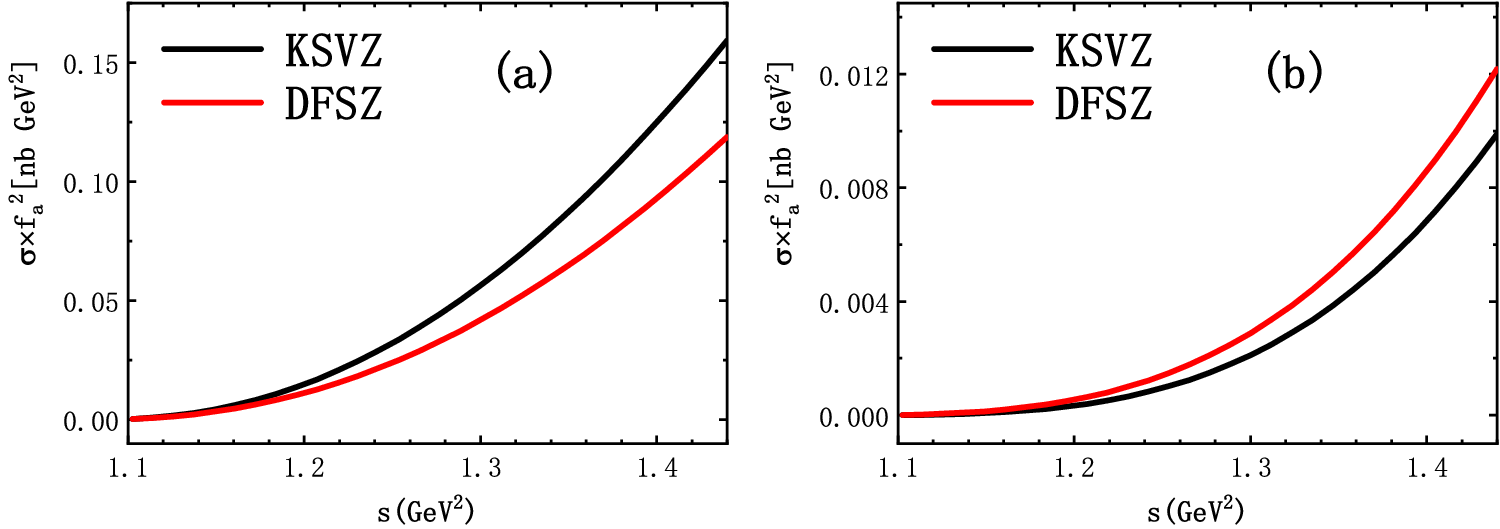}
\caption{Total cross sections for $ep\rightarrow epa$ (a) and $en\rightarrow ena$ (b) with $m_a=0.1$~GeV.}
\label{Figure:tcep}
\end{figure}

We first focus on the electron case. The differential cross sections for the proton target with respect to the axion energy $E_a$ at two different axion masses with $m_a\sim 0$ and $m_a=100$~MeV are given for both the KSVZ and DFSZ models in Fig.~\ref{Figure:tcepEa}. For the model-dependent axion parameters in the DFSZ case, we fix $E/N=44/3$ for the axion two-photon coupling and $\sin^2\beta=1/2$ for the preexisting axion-quark coupling, both of which values will also be taken in Fig.~\ref{Figure:dcepangle} for the differential cross sections of the $ep\to epa$ process with respect to the angles $\varphi^\ast, \theta^\ast$, and $\theta$. In the latter figure, we show the results for two axion masses at $m_a\sim 0$ and $m_a=100$~MeV as well.  
For illustration, the total cross sections of the $ep\to epa$ reaction as a function of $s$ with $m_a=100$~MeV are shown in Fig.~\ref{Figure:tcep} for both the KSVZ and DFSZ cases. Similar magnitudes and trends of the curves in the $ep\to epa$ process are observed for the two different axion models. 

\begin{figure}[htbp]
\centering
\includegraphics[width=0.75\textwidth]{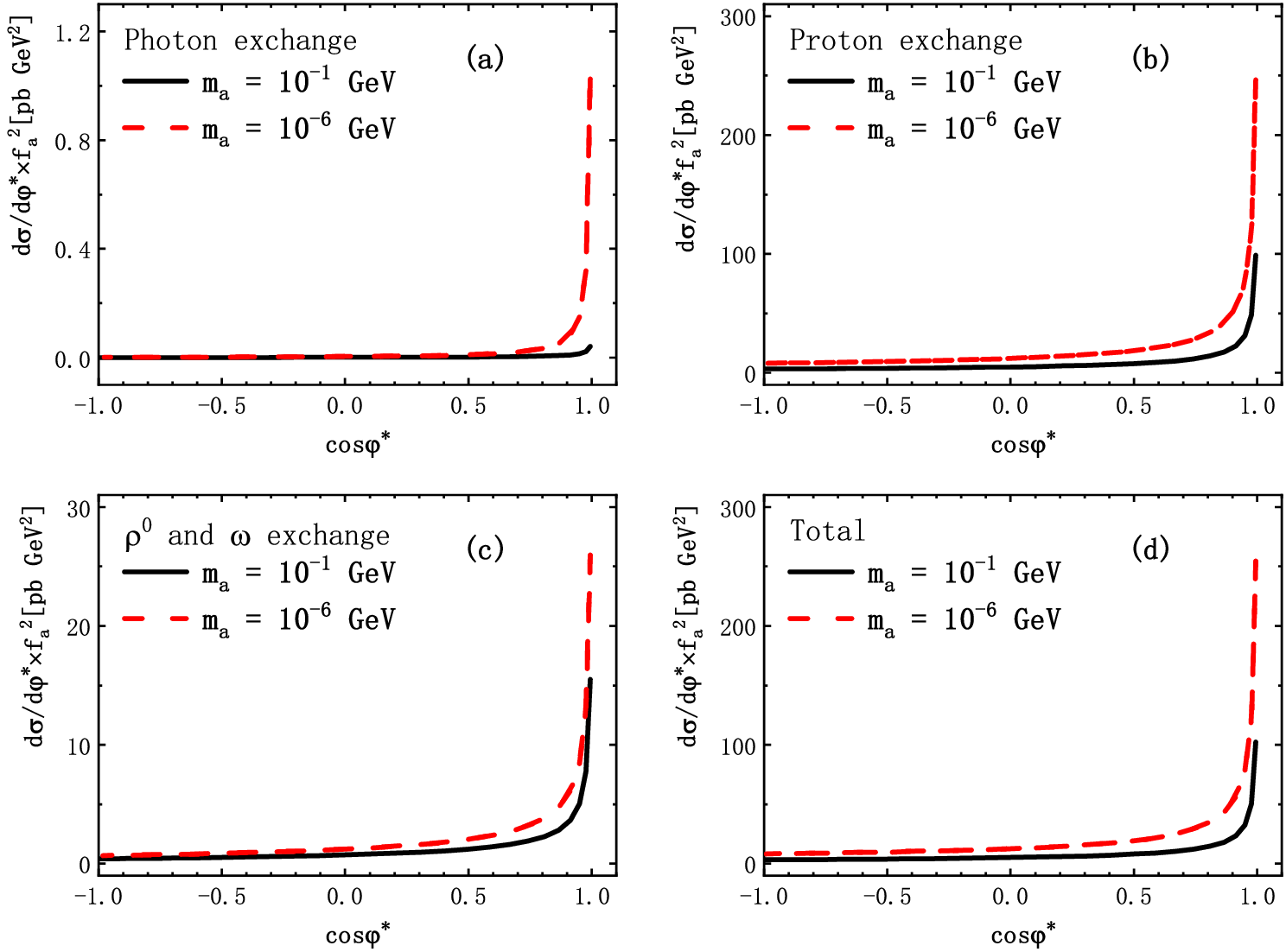}
\caption{Differential cross section $\md\sigma/\md\varphi^*$ with respect to $\cos{\varphi^*}$ for $ep\rightarrow epa$ in the case of the KSVZ model, evaluated for different exchange mechanisms: photon exchange (a), proton exchange (b), combined $\rho^0$ and $\omega$ exchanges (c), and the full amplitude including all exchanges (d). }
\label{Figure:dcepphistar}
\end{figure}

\begin{figure}[htbp]
\centering
\includegraphics[width=0.75\textwidth]{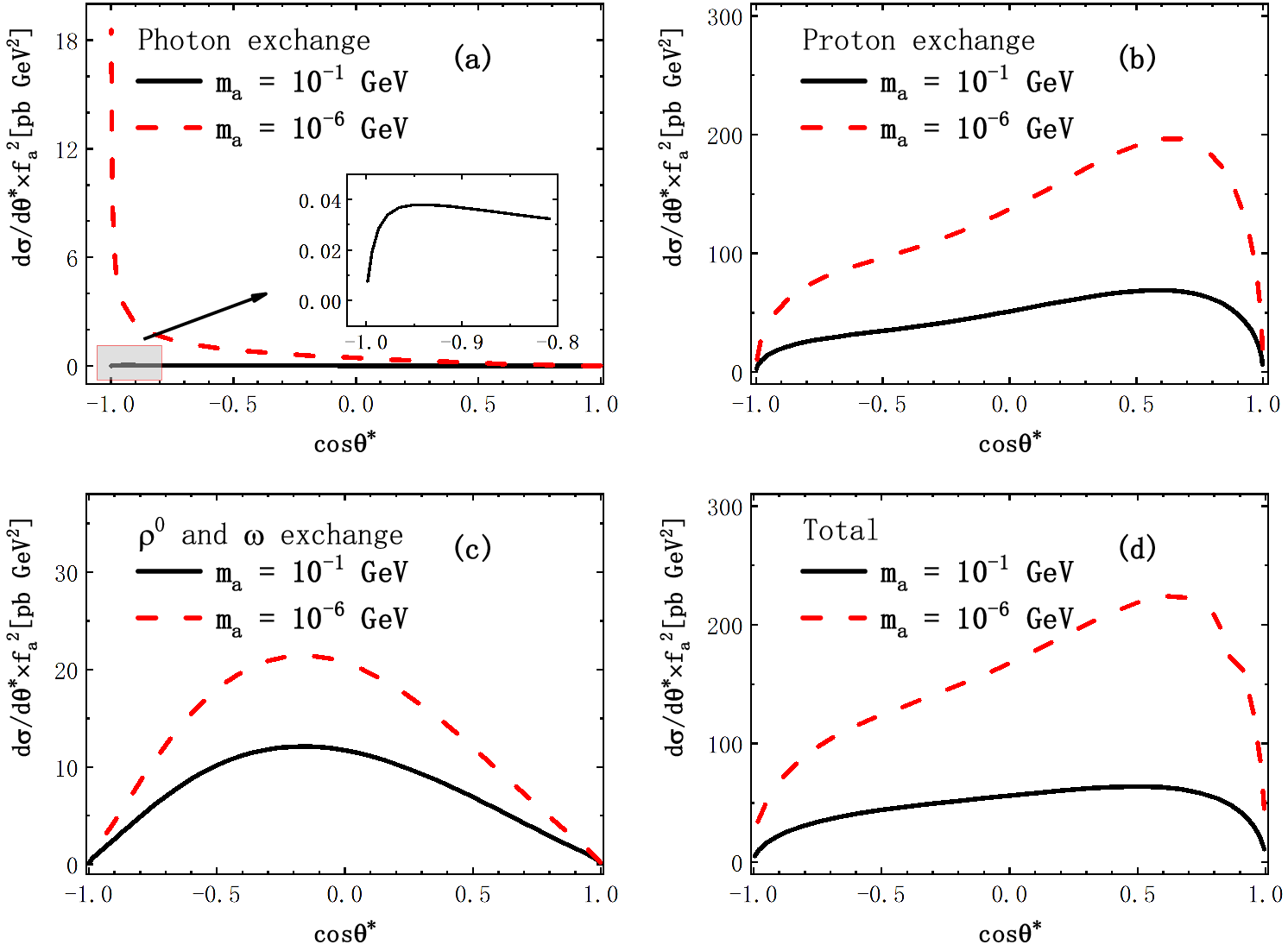}
\caption{Differential cross section $\md\sigma/\md\theta^*$ with respect to $\cos{\theta^*}$ for $ep\rightarrow epa$ in the case of the KSVZ model, evaluated for different exchange mechanisms: photon exchange (a), proton exchange (b), combined $\rho^0$ and $\omega$ exchanges (c), and the full amplitude including all exchanges (d).}
\label{Figure:dcepthetastar}
\end{figure}

\begin{figure}[htbp]
\centering
\includegraphics[width=0.75\textwidth]{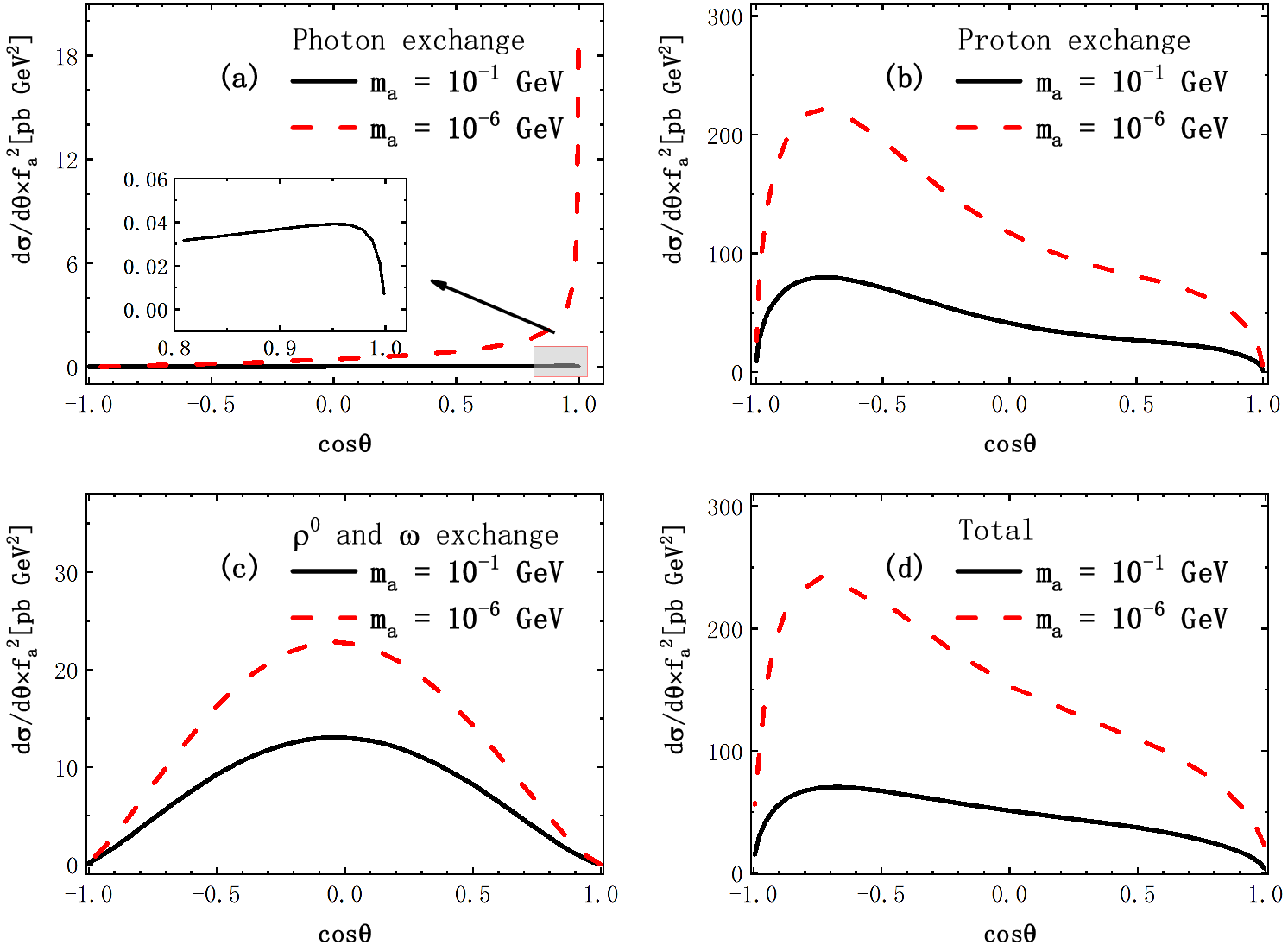}
\caption{Differential cross section $\md\sigma/\md\theta$ with respect to $\cos{\theta}$ for $ep\rightarrow epa$ in the case of the KSVZ model, evaluated for different exchange mechanisms: photon exchange (a), proton exchange (b), combined $\rho^0$ and $\omega$ exchanges (c), and the full amplitude including all exchanges (d).}
\label{Figure:dceptheta}
\end{figure}

\begin{figure}[htbp]
\centering
\includegraphics[width=0.75\textwidth]{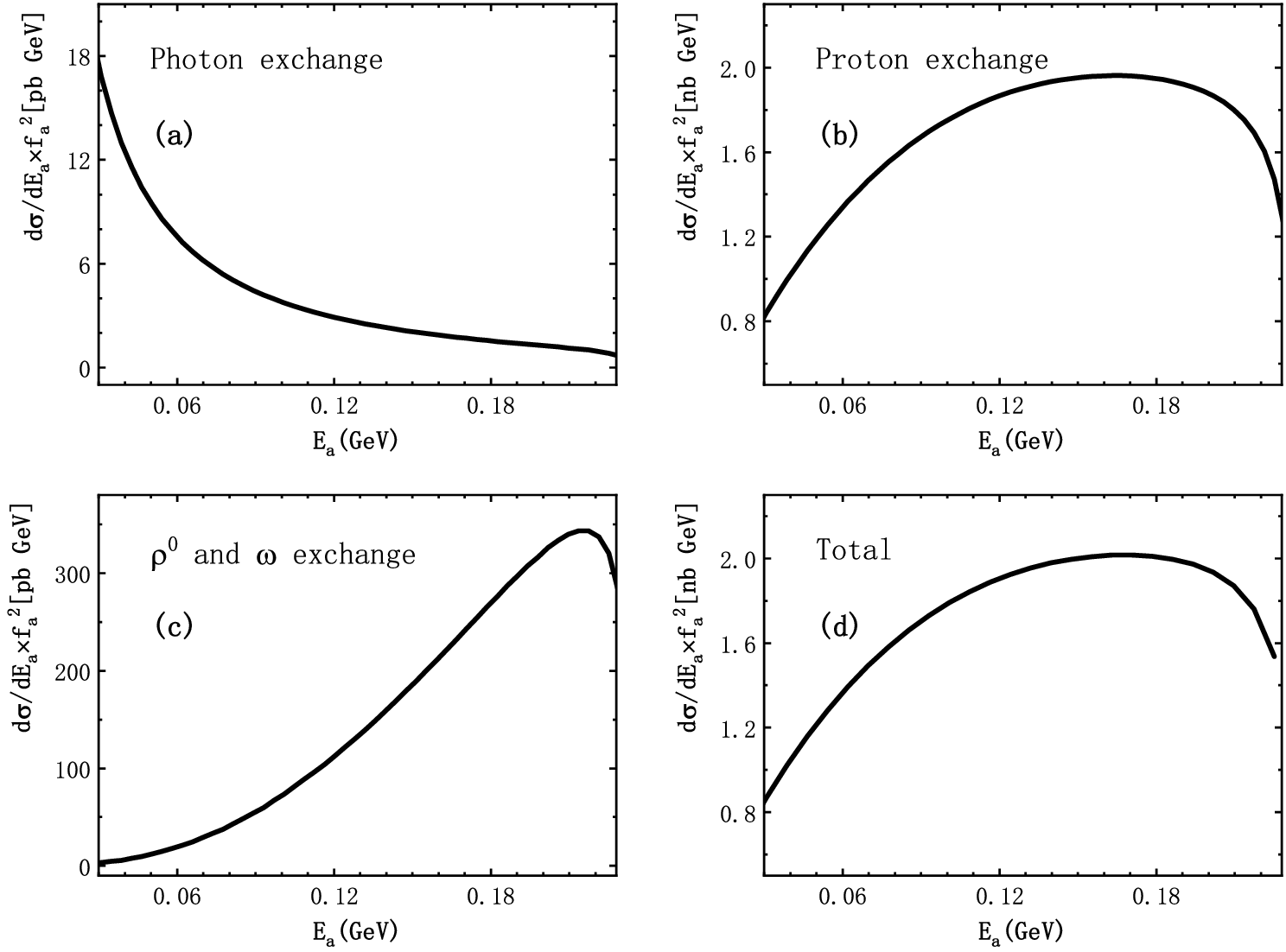}
\caption{Differential cross section $\md\sigma/\md E_a$ with respect to $E_{a}$ for $ep\rightarrow epa$ at $m_a=10^{-6}$~GeV in the case of the KSVZ model, evaluated for different exchange mechanisms: photon exchange (a), proton exchange (b), combined $\rho^0$ and $\omega$ exchanges (c), and the full amplitude including all exchanges (d).}
\label{Figure:dcepEa}
\end{figure}

In order to analyze the effects of different mechanisms illustrated in the Feynman diagrams of Fig.~\ref{Figure:ep-epa-tree}, we separately calculate the differential distributions contributed from each individual exchange of different particles, including photon, nucleon, and vector resonances, namely the amplitude $\mathcal{M}$ in Eq.~\eqref{eq.defcs} will be correspondingly taken as $\mathcal{M}_{a1}$, $\mathcal{M}_{b}+\mathcal{M}_{c}$, $\mathcal{M}_{a2}+\mathcal{M}_{a3}$ in order. The resulting curves are shown in Figs.~\ref{Figure:dcepphistar}, \ref{Figure:dcepthetastar}, \ref{Figure:dceptheta}, and \ref{Figure:dcepEa} for the distributions of $\phi^\ast$, $\theta^\ast$, $\theta$, and $E_a$, respectively. In order not to overload the figures, we only show the results from the KSVZ model. In this case, the proton exchange via the $g_{aNN}$ vertices gives the dominant contributions in the four types of distributions in Figs.~\ref{Figure:dcepphistar}-\ref{Figure:dcepEa}. The contributions from the exchanges of vector resonances and photon are respectively around 1 and 2 orders of magnitude smaller than that of the nucleon exchange. It is pointed out that the magnitudes from different mechanisms are subject to the axion model setups. However, the line shapes of the curves corresponding to different types of particle exchanges remain robust. For the distributions with respect to $\varphi^\ast$, different types of particle exchanges lead to rather similar line shapes, i.e., they all give prominent contributions around the region with $\varphi^\ast\to 0$, as shown in Fig.~\ref{Figure:dcepphistar}, implying that the $\varphi^\ast$ distribution is not a proper quantity to distinguish different microscopic mechanisms in the $e p\to e p a$ process. In contrast, the distributions with respect to $\theta^\ast$, $\theta$, and $E_a$ are found to be able to provide useful quantities to distinguish different mechanisms among the exchanges of photon, nucleon, and vector resonances, since the three types of particle exchanges lead to apparently distinct line shapes as shown in Figs.~\ref{Figure:dcepthetastar}, \ref{Figure:dceptheta}, and \ref{Figure:dcepEa}. We verify that by taking the inputs of the DFSZ model as explained previously, qualitatively similar conclusions can be obtained. 
It is pointed out that the automation toolkit, \texttt{MadGraph}~\cite{Alwall:2014hca}, has been exploited to compute the various differential cross sections for the purpose of cross-checking the results presented in this work. The model description of the electron-proton-axion.fr file can be acquired via the link in~\cite{electron-proton-axion-fr}. We find that the outputs of \texttt{MadGraph} with separate contributions from the exchanges of nucleon and vector mesons are in excellent agreement with our own calculations, while \texttt{MadGraph}'s result for the photon exchange shows a large uncertainty that is more than 30\%. 

\begin{figure}[htbp]
\centering
\includegraphics[width=0.75\textwidth]{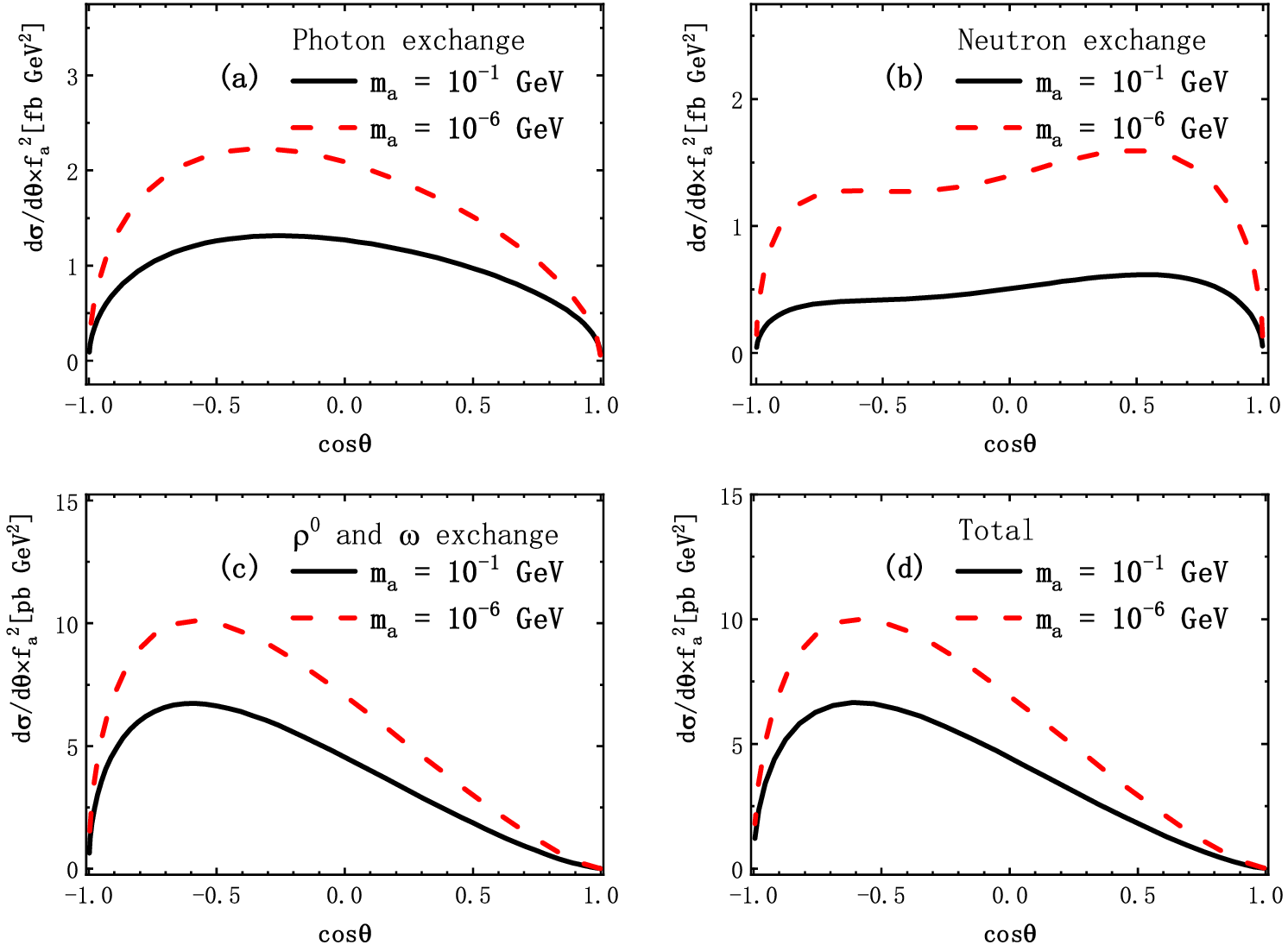}
\caption{Same as Fig.~\ref{Figure:dceptheta} but for $en\rightarrow ena$.  }
\label{Figure:dcentheta}
\end{figure}

\begin{figure}[htbp]
 \centering
 \includegraphics[width=0.75\textwidth]{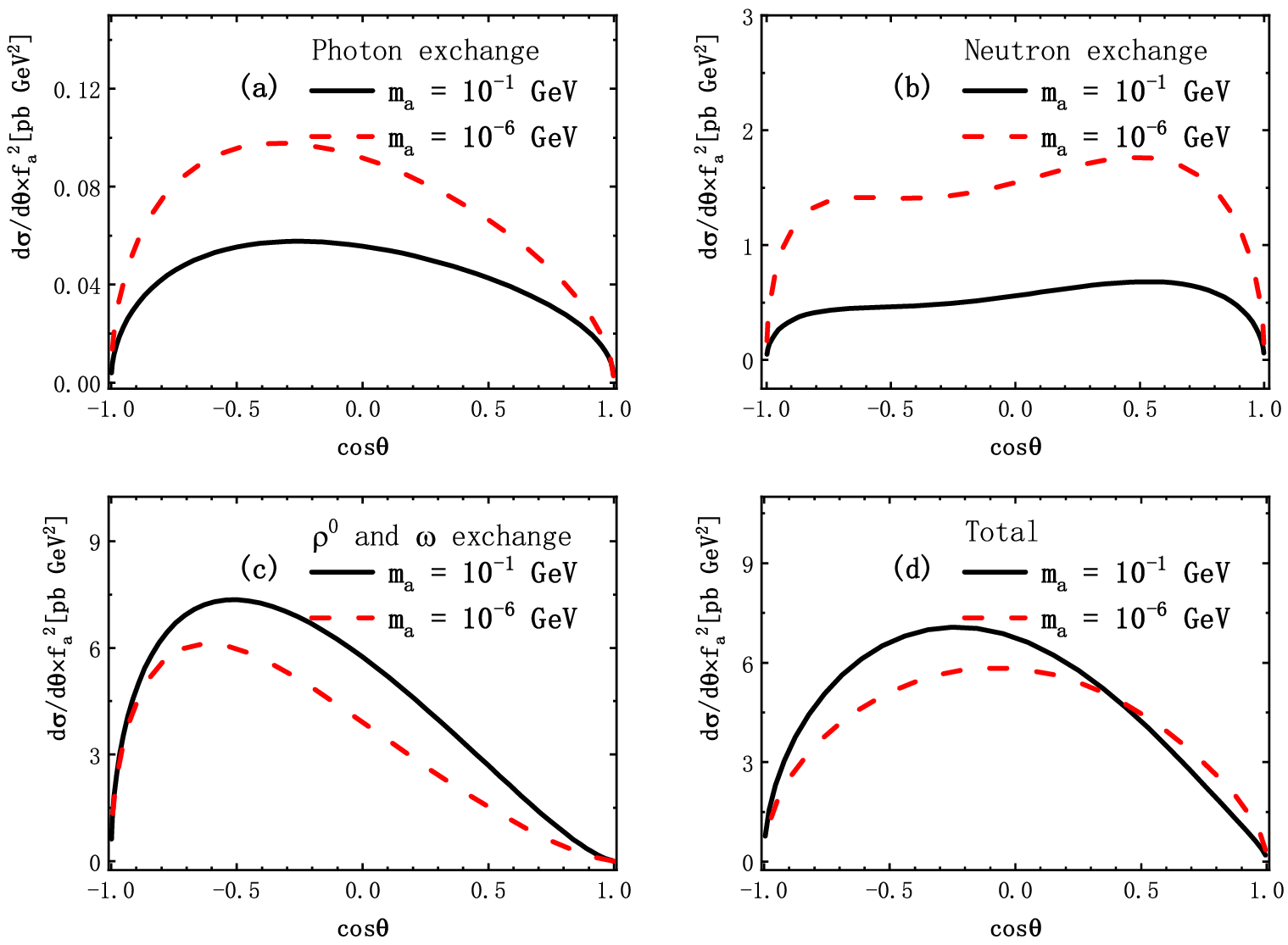}
 \caption{Same as Fig.~\ref{Figure:dceptheta} but for $en\rightarrow ena$ in the case of the DFSZ axion model. }
 \label{Figure:dcenthetaDFSZ}
\end{figure}

\begin{figure}[htbp]
\centering
\includegraphics[width=0.75\textwidth]{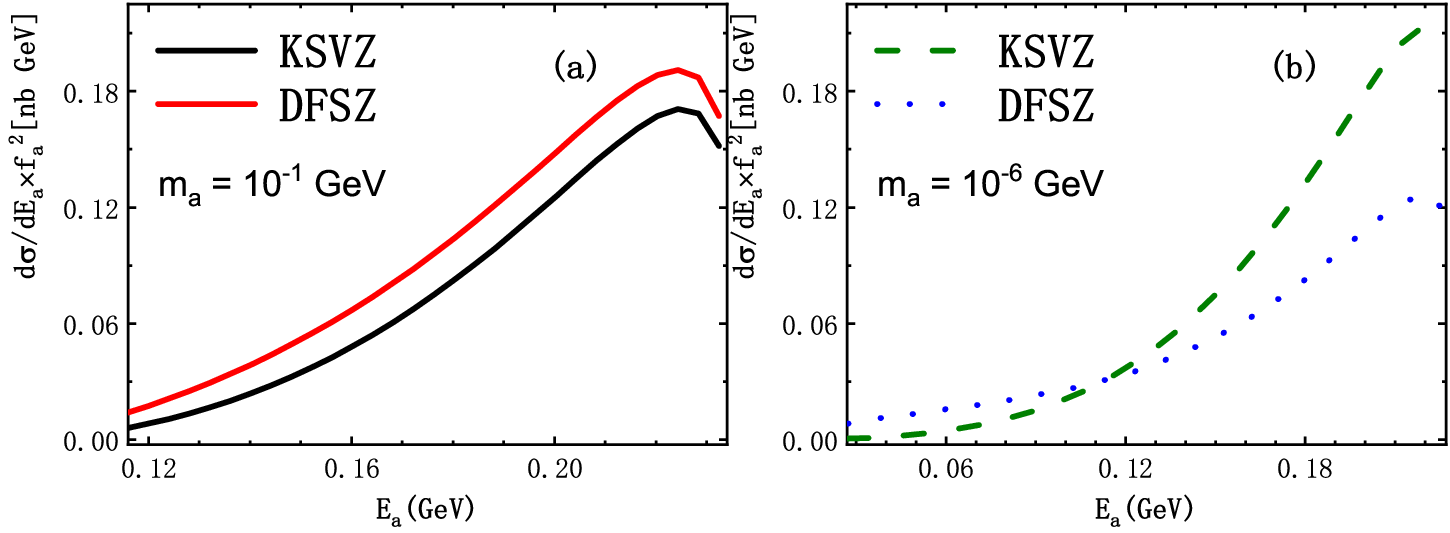}
\caption{Same as Fig.~\ref{Figure:tcepEa} but for $en\rightarrow ena$.}
\label{Figure:dcenEa}
\end{figure}

\begin{figure}[htbp]
\centering
\includegraphics[width=1\textwidth]{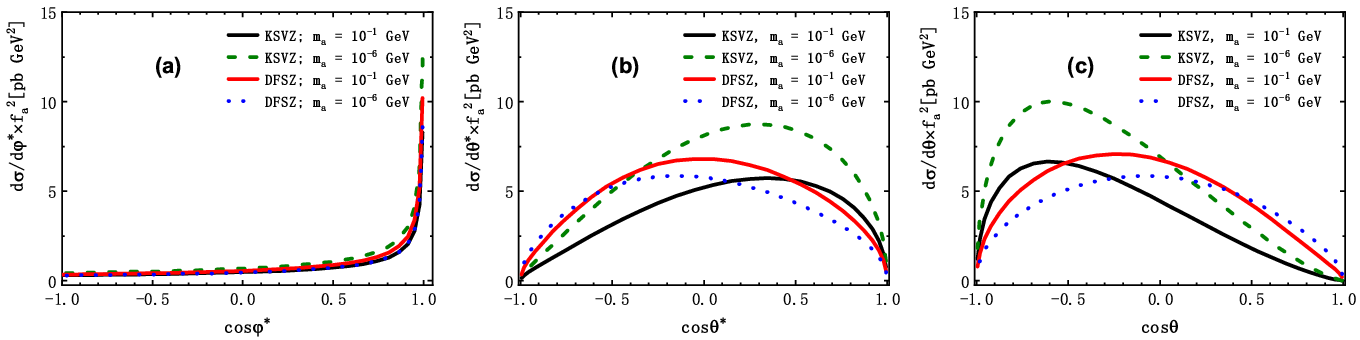}
\caption{Same as Fig.~\ref{Figure:dcepangle} but for $en\rightarrow ena$.}
\label{Figure:dcenangle}
\end{figure}

Then, we explore the phenomenological results for ALP production in the electron-neutron case. A striking difference from the proton target case is that the exchanges of vector mesons dominate the $e n \to e n a$ process. To visualize this phenomenon, the differential cross sections with respect to $\cos\theta$ are taken as an example in Fig.~\ref{Figure:dcentheta}, where we separately analyze results from different types of particle exchanges. The magnitudes of the vector meson exchanges are more than 3 orders larger than those from the exchanges of photons and nucleons in the KSVZ case; the reason is that the axion-neutron coupling $g_{ann}$ happens to be rather small in the KSVZ model. In the DFSZ case, the model dependent part of the axion-neutron coupling becomes sizable and it leads to similar magnitudes with the vector meson exchanges, as shown in Fig.~\ref{Figure:dcenthetaDFSZ}. The differential cross sections by taking the full amplitudes with respect to $E_a$ for the axion masses at $m_a\sim 0$ and 100~MeV are given in Fig.~\ref{Figure:dcenEa}, and the distributions with respect to $\varphi^\ast$, $\theta^\ast$, and $\theta$ are illustrated in Fig.~\ref{Figure:dcenangle}. Compared with the results of the proton target case in Figs.~\ref{Figure:tcepEa}, \ref{Figure:dcepangle} and ~\ref{Figure:tcep}, the magnitudes of various differential cross sections in the neutron channel are more than 1 order smaller. 
By taking the ansatz of $\sigma_{\ell-{\rm nucleus}} = Z_p^2 \sigma_{\ell-{\rm proton}}+  Z_n^2\sigma_{\ell-{\rm neutron}}$ discussed previously and the electron-nucleon cross sections illustrated in Fig.~\ref{Figure:tcep}, one can give a rough estimation of the electron-nucleus total cross sections for various targets in the low energies. For example, when taking tungsten ($W$) as the target, i.e., $Z_p=74$ and $Z_n=110$, the corresponding cross section of $e+W\to e + W +a$ with the CM energy of the incoming electron at $0.3$~GeV (corresponding to the energy of the initial $eN$ system at $\sqrt{s}=1.2$~GeV) would be at a level of around $0.8\sim 1.0$ $\mu$b$\cdot {\rm GeV^2}/f_a^2$. Following the same recipe, we also estimate the cross section of the axion photoproduction off the tungsten nucleus target by taking the results of axion photoproduction off the nucleon, i.e., the $\gamma + N \to N + a$, from Ref.~\cite{Cao:2024cym} as inputs. The cross section of $\gamma + W \to W +a$ with the CM energy of the incoming photon at $0.3$~GeV (corresponding to the energy of the initial $\gamma N$ system at $\sqrt{s}=1.2$~GeV) lies in the range of $190\sim 360$ $\mu$b$\cdot $ GeV$^2/f_a^2$, which is 2 orders of magnitude larger than that of the $e+W\to e + W +a$ process. The comparison of the $e + A \to e + A +a$ and $\gamma + A \to A +a$ cross sections could provide relevant criteria to discriminate the relative importance of the two mechanisms, leveraging track-length information from different materials~\cite{Darme:2020sjf}.

\begin{figure}[t]
\centering
\includegraphics[width=0.75\textwidth]{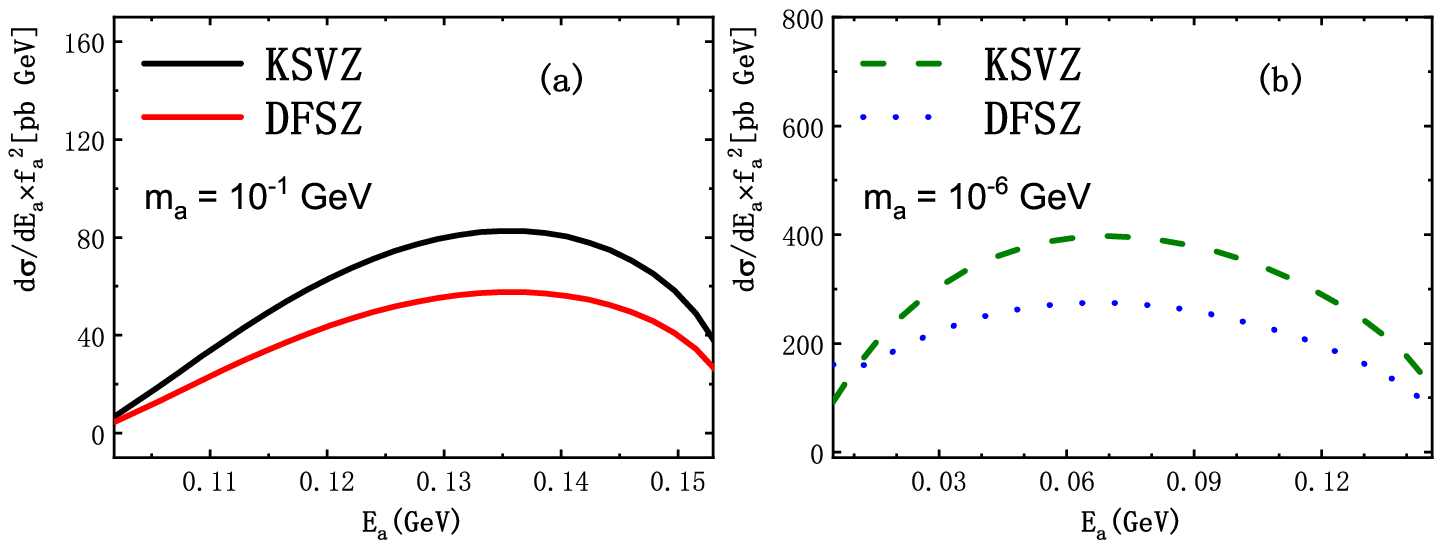}
\caption{Same as Fig.~\ref{Figure:tcepEa} but for $\mu p\rightarrow \mu pa$. }
\label{Figure:tcmupEa}
\end{figure}

\begin{figure}[htbp]
\centering
\includegraphics[width=0.75\textwidth]{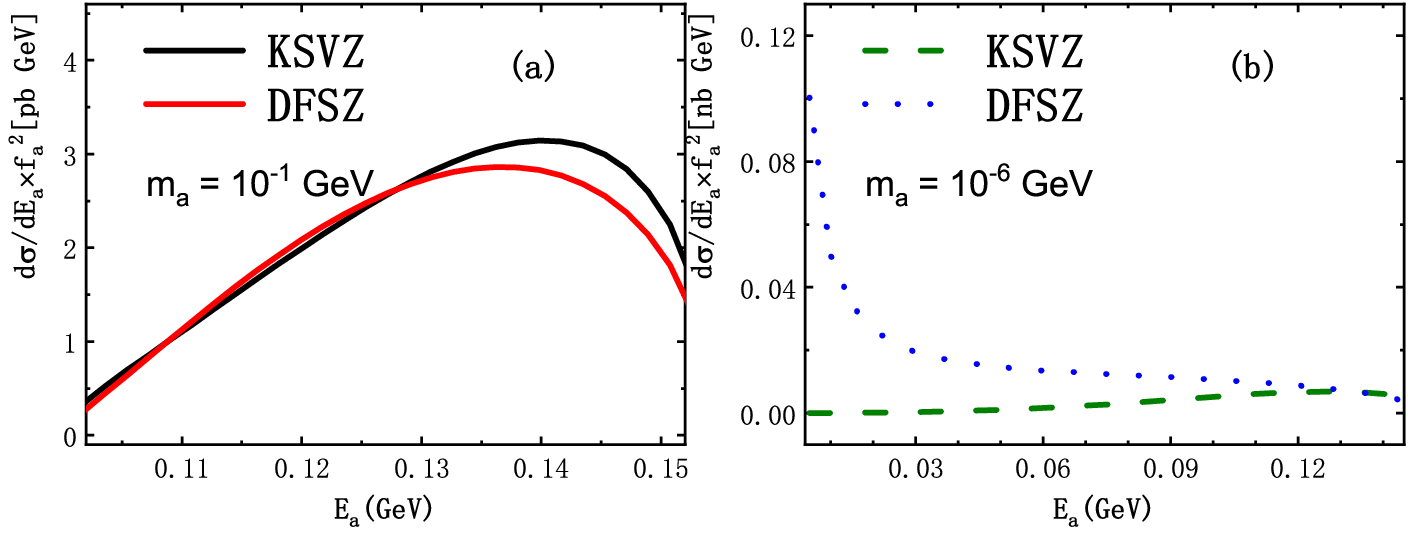}
\caption{Same as Fig.~\ref{Figure:tcepEa} but for $\mu n\rightarrow \mu na$.}
\label{Figure:dcmunEa}
\end{figure}

\begin{figure}[t]
\centering
\includegraphics[width=1\textwidth]{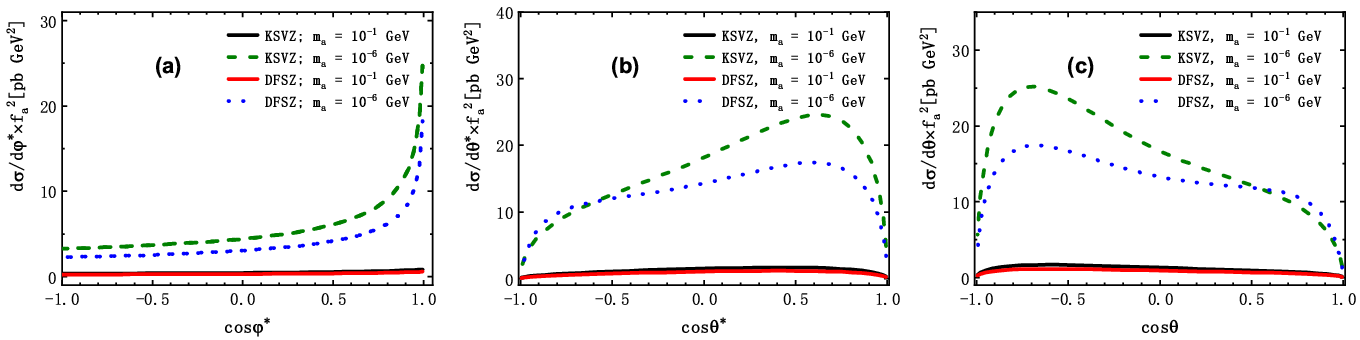}
\caption{Same as Fig.~\ref{Figure:dcepangle} but for $\mu p\rightarrow \mu pa$. }
\label{Figure:dcmupangle}
\end{figure}

\begin{figure}[htbp]
\centering
\includegraphics[width=1\textwidth]{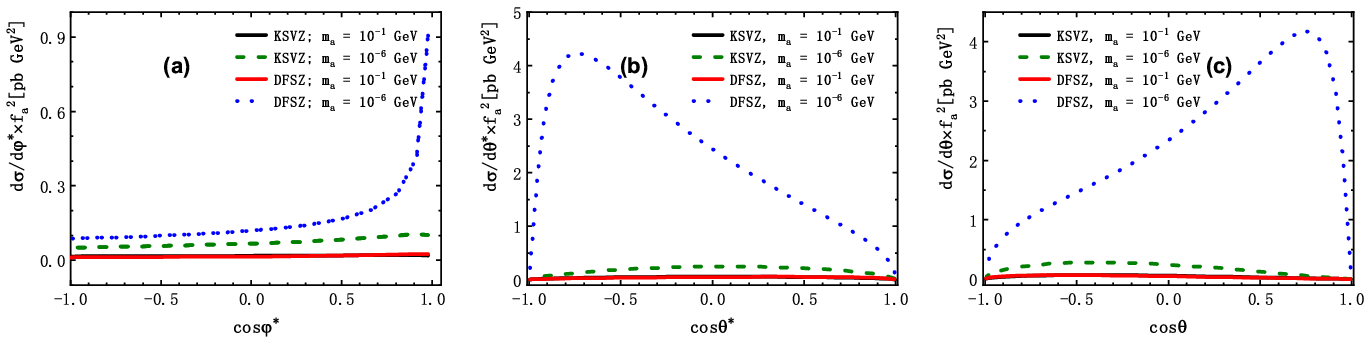}
\caption{Same as Fig.~\ref{Figure:dcepangle} but for $\mu n\rightarrow \mu na$.}
\label{Figure:dcmunangle}
\end{figure}

Next, we briefly discuss the results when replacing the incoming electron with muon. Generally speaking, the trends of various curves of the differential cross sections are similar to those in the electron case. As expected, the magnitudes of the $\mu N \to \mu N a$ cross sections are much smaller than those of the electron case in the low-energy regions, due to the phase space suppression. To be specific, we illustrate the $E_a$ distributions in Figs.~\ref{Figure:tcmupEa} and \ref{Figure:dcmunEa} for $\mu p\to \mu p a$ and $\mu n\to \mu n a$, respectively, which are obtained by taking the CM energy at $\sqrt{s}=1.2$~GeV. With the same CM energy, the differential cross sections with respect to the cosines of various angles by taking the full amplitudes are given in Figs.~\ref{Figure:dcmupangle} and ~\ref{Figure:dcmunangle} for $\mu p\to \mu p a$ and $\mu n\to \mu n a$ processes, respectively. To avoid overcrowding the discussion with excessive figures, we omit showing the various differential cross sections by separately including the individual exchanges of the photon, nucleon, and vector mesons, since they roughly exhibit behaviors qualitatively similar to the electron situation.

\section{Summary and conclusions}\label{sec:sum}

We have carried out a thorough study of the ALP production from low-energy lepton-nucleon scattering processes, i.e., $\ell N\to \ell N a$, with $N$ the proton and neutron. Three different types of interaction vertices, including the ALP-nucleon-nucleon, ALP-photon-photon and ALP-photon-vector meson ones,  are simultaneously taken into account in our study. Chiral effective field theory is employed to construct these three types of interaction operators. Extensive inputs from the lattice QCD and hadron phenomenological studies are taken to constrain the unknown hadron-related couplings, so that we are only left with the unknown ALP parameters, e.g., $f_a$ and $m_a$ in the KSVZ case and additional ones with $\tan\beta$ and $E/N$ in the DFSZ case. Since $f_a$ appears as a global factor in all the amplitudes, we have multiplied the (differential) cross sections by $f_a^2$ in the phenomenological discussions. By taking two different ALP masses at $m_a\sim 0$ and $m_a=100$~MeV, our study shows that the differential and total cross sections are mildly affected by the ALP masses. The relative strength of different ALP interaction vertices is found to be rather different in the proton and neutron channels. Generally speaking, the vector resonance exchanges play important roles in the $\ell n\to \ell n a$ process regardless of the KSVZ or DFSZ models, while in the $\ell p\to \ell p a$ process, the dominant contributions are given by the nucleon exchanges via the ALP-proton-proton coupling, and the exchanges of vector meson resonances and photons can be around 1 or 2 orders smaller, depending on different axion model setups. Not only can the three different types of ALP couplings lead to different magnitudes, but also they cause rather distinct line shapes with respect to the various angles and ALP energy.  
It is hoped that the sophisticated low-energy chiral amplitudes for the $\ell N\to \ell N a$ processes calculated in this work can provide useful inputs and theoretical constraints for related future studies.

\section*{Acknowledgements}
This work is funded in part by the Natural Science Foundation of China (NSFC) under Grants Nos.~12475078, 12150013, 11975090 and 12047503; the Postdoctoral Fellowship Program of the China Postdoctoral Science Foundation under Grants Nos.~GZC20232773 and ~2023M74360; and the Science Foundation of Hebei Normal University with contract No.~L2023B09. 

\section*{Data Availability}

The model description in FeynRules (the electron-proton-axion.fr file) is openly available~\cite{electron-proton-axion-fr}.

\bibliography{eNeNa}
\bibliographystyle{apsrev4-2}

\end{document}